\documentclass[%
 preprint,        
 amsmath,amssymb,
 aps,
prstab,
]{revtex4-2}

\usepackage{graphicx}
\usepackage{dcolumn}
\usepackage{bm}
\usepackage[mathlines]{lineno}

\begin{document}


\preprint{APS/123-QED}

\title{Characteristics of Dark Current from a S-band RF Gun with Exchangeable Photocathode System}

\author{F.Jackson}
\email{frank.jackson@stfc.ac.uk} %
\author{L. S. Cowie}%
\author{M. Furmaniak}%
\author{P. Goudket}%
\author{B. L. Militsyn}%
\author{J. W. McKenzie}%
\author{T. C. Q.  Noakes}%

\affiliation{%
 UKRI/STFC Daresbury Laboratory\\
}%



\begin{abstract}
Field emission or dark current from RF electron guns is problematic for accelerators for several reasons, and can be more difficult to predict, measure and understand, compared to the main beam. The work presented here is a study of the dark current emitted from the CLARA 3GHz 2.5 cell electron gun with exchangeable photocathode, which is a type of gun used for single pass high quality low emittance beams for uses such as free electron lasers. The main purpose of these studies was to identify sources of dark current emission within the gun, by examining the dark current distribution and patterns on scintillating screens.  It is shown that the features of the patterns can be reproduced qualitatively by finite element simulations. Thus, certain surfaces within the gun and cathode apparatus which contribute most significantly to problematic dark current can be identified, and further development and preparation of the guns and their cathodes can be informed to minimise dark current.
\end{abstract}

\maketitle


\section{\label{sec:level1}Introduction}

Field emission of electrons in RF guns of the type used to produce high quality beams such as for free electron lasers (FELs) has the potential to cause many different problems related to machine protection. The capture and propagation of this field emission is known as `dark current' and may affect different parts of an accelerator. In FELs for example, gun dark current transported to FEL undulators may cause undulator damage~[\onlinecite{frohthesis}].  

CLARA (compact linear accelerator for research and applications) at Daresbury laboratory was originally designed as a FEL test facility~[\onlinecite{Clarke_2014}], and includes a high gradient RF electron gun similar to that used in existing x-ray FELs (XFELs). A description of the first operational CLARA gun, and guns at the European XFEL and linac 
coherent light source (LCLS) facilities can be found within the references~[\onlinecite{PhysRevAccelBeams.23.044801}]~[\onlinecite{brinker:ipac16-tuoca03}]~[\onlinecite{PhysRevSTAB.11.030703}].  Dark current from the CLARA gun has been observed since first commissioning and has been of concern as the machine has developed. For example, the background level caused by gun dark current has been recorded in some non-FEL experiments and applications (electron diffraction studies, very high electron energy for radiotherapy investigations)~[\onlinecite{rudge:ipac15-tupwi017}]~[\onlinecite{agnesthesis}]. Most recently the level of dark current, following a gun upgrade, has increased significantly to a level where the first beam diagnostic screen can be dominated by dark current, and concerns remain about its effect on cathode performance, damage to diagnostic devices like charged couple devices, discolouration of beamline viewports, and even potentially effects on the optical components of the photoinjector laser transport line. 

The understanding and mitigation of dark current from the CLARA gun, and other similar guns, presents several difficulties. The field emission may be generated from several different parts of the inner surface of the gun, and the field emission particles have wide ranging kinematic properties compared to the core electron beam, in particular a large spread of energies. The physical and chemical properties of the photocathode and other gun parts, particularly in the cathode area, have a strong influence on the dark current. However the effects are not well understood, and difficult to measure to a high degree of precision. Disentangling the effects on the macroscopic features of dark current, typically observed on scintillating beam diagnostics screens, is a complex problem. There have been several investigations of these features in different electron guns worldwide, some similar to the CLARA gun. For the LCLS gun an analytical model has provided a representation of the dark current screen patterns~[\onlinecite{4441062}]. The advanced photon injector experiment (APEX) gun at Lawrence Berkeley national laboratory (LBNL) has investigated the dark current screen appearance and associated it to emission points in the cathode area ~[\onlinecite{PhysRevSTAB.18.013401}]. Detailed simulations have been used for some guns to model dark current emission and transport to screens: at the Argonne wakefield accelerator facility (AWA) the macroscopic distribution of surface microscopic field emitters was deduced from screen measurements using these simulations and achieved a high degree of agreement with surface scanning electron microscopy measurements~[\onlinecite{PhysRevLett.117.084801}]; a similar simulation has been used to model dark current transport at Polish free electron laser (POLFEL)~[\onlinecite{szymczyk:fel17-wep021}]; recently a detailed study of dark current screen patterns at the photoinjector test facility at DESY, Zeuthen site (PITZ) gun has successfully reproduced the observed features in simulation and associated them with surface locations of the cathode~[\onlinecite{SHU2021165546}]. Modelling of the RF electromagnetic (EM) field and consideration of microscopic emission points, particularly on the surfaces of the cathode area of the guns, has been a key element of the above studies. 

For the CLARA gun, particle-in-cell (PIC) simulations have been developed using CST Studio~[\onlinecite{cst}], to understand the appearance of dark current screen observations, particularly following the gun upgrade. In the following paper we describe the simulation of dark current generation and transport in the CLARA gun and injector, and a comparison of its features to screen observations. It is shown that components of the cathode area can be identified as emission sources using these simulations. The results can be used to inform future gun development to mitigate dark current.   

\section{\label{sec:level1}THE CLARA INJECTOR AND GUN}
A recent description of the CLARA accelerator is given in~[\onlinecite{PhysRevAccelBeams.23.044801}]. The CLARA gun and photoinjector have been in operation at Daresbury laboratory since 2013. The gun is a 2.5 cell copper cavity designed to reach a maximum field of 100 MV/m and produce a beam of several MeV energy in single bunches operating at 10 Hz repetition rate. In the initial installation the gun copper backplate served as a bulk photocathode and the electron beam was generated with a laser of wavelength 266 nm. In 2019 the gun was upgraded by replacing the backplate with a cathode exchange mechanism, and since then hybrid copper cathodes of different properties have been commissioned~[\onlinecite{noakes:ipac2023-tupa032}].

\begin{figure*}[!htb]
\includegraphics[width=0.75\textwidth]{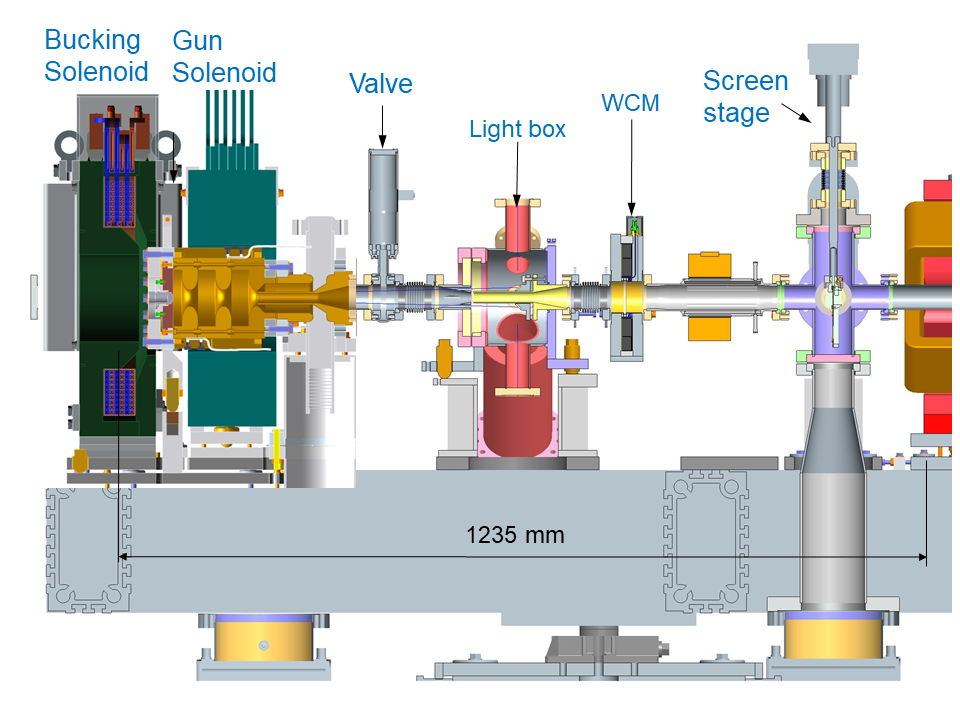}
\caption{\label{fig:gun1} The upgraded 2.5 cell CLARA gun beamline, including the gun and its solenoids, the light box, the wall current monitor (WCM) and the first screen. The gun cathode plug can be inserted into the back of the gun (the load-lock mechanism is not included in the diagram). The focussing solenoids surrounding the gun (main and bucking) are displayed.}
\end{figure*}

\begin{figure}[!htb]
\centering
\includegraphics[width=0.5\textwidth]{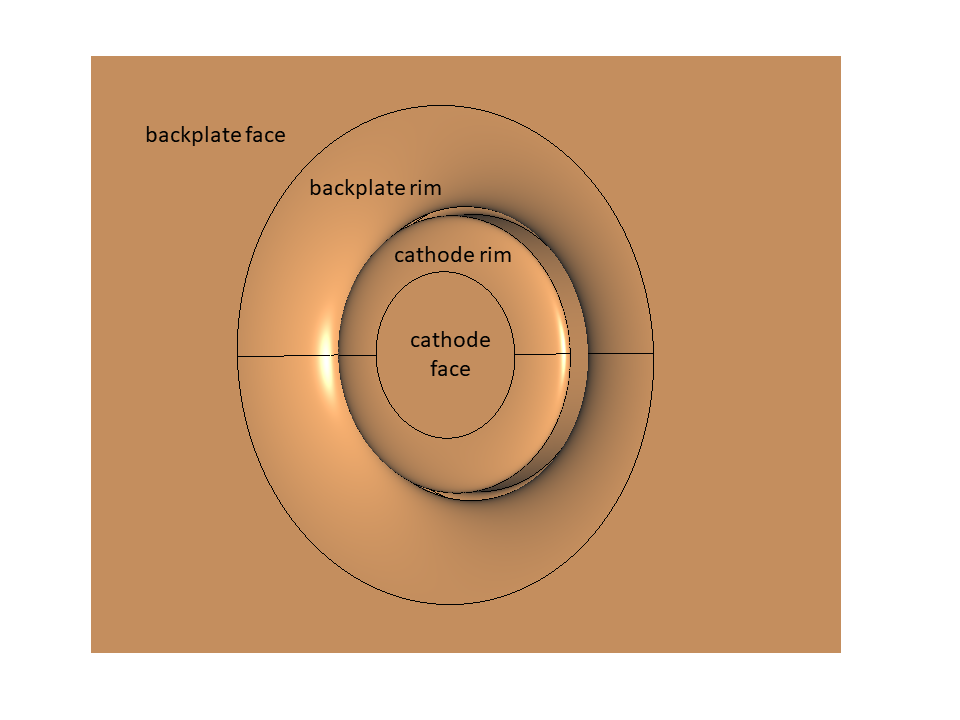}
\caption{\label{fig:gun2} Close up view of the CLARA gun cathode region, consisisting of the retractable cathode insert(plug) and opening in the gun backplate in which it sits.}
\end{figure}

Figs.~\ref{fig:gun1} and~\ref{fig:gun2} show the upgraded CLARA gun and mechanical design of the gun region. Compared to the first installation, the gun backplate was redesigned with a cathode exchange mechanism. The gun incorporates a main solenoid magnet to allow beam focussing, and a ‘bucking’ solenoid to cancel the solenoid field on the surface of the cathode. The design of the cathode exchange mechanism was influenced by a project to develop an alternative high repetition rate gun for CLARA (up to 400 Hz). This development sought to optimise performance of the gun, through control of the electromagnetic field by shaping of inner gun surfaces including those of the cathode areas. A key element of this design was the elliptical design of cavity irises and the profile of the cathode insert (plug). The cathode plug profile has a flat central section of 6 mm diameter and an elliptical edge (rim). The gun electric field is enhanced at the cathode edges by the plug opening. In the design the ellipticity was optimised to maximise the ratio of the peak field in the centre of the cathode to that on the edges. An elliptical shape falling back to 0.6 mm from the cathode face was found to be the optimum plug shape, achieving a ratio of around 0.9. In other words, even when optimised, the peak field on the cathode edges was still slightly larger than the field in the centre (as can be seen in Fig.  ~\ref{fig:cst2}).


The CLARA machine and injector beamline is described elsewhere~[\onlinecite{PhysRevAccelBeams.23.044801}]; the injector line consists of the gun and a beampipe approximately 1m long containing a light box through which the photoinjector laser is directed onto the cathode, a wall current monitor (WCM) to measure beam charge, and a scintillation (YAG) screen on which the beam and the dark current emerging from the gun are observed. Immediately after the screen is the first main accelerating cavity. 

\section{\label{sec:level1}CLARA DARK CURRENT SIMULATION OUTLINE}

Finite element method (FEM) simulations of the dark current generation and transport in the CLARA injector have been developed using CST. Both the electromagnetic field in the gun and the dark current are simulated within CST, based on FEM methods. The solenoid magnetic field is simulated in an external code OPERA~[\onlinecite{opera}] and imported into CST. This method provides a full 3-D model of the situation. It was first used for simulations of dark current in the CLARA gun with its original bulk photocathode and was able to provide a level of agreement with experimental observations~[\onlinecite{jackson:ipac17-wepab088}]. When the CLARA gun was upgraded the CST simulations were modified to reflect the changes, and the complete model used here includes the gun with new cathode geometry, the coupler geometry, and the beam pipe extending through the light box and its narrower aperture, up to the first beam diagnostic screen. The beampipe was modelled as a 35 mm diameter cylinder, except at the gun RF coupler and light box assembly where the apertures were modelled as 15 mm and 17 mm diameter respectively.  

\begin{figure}[!htb]
\centering
\includegraphics[width=0.5\textwidth]{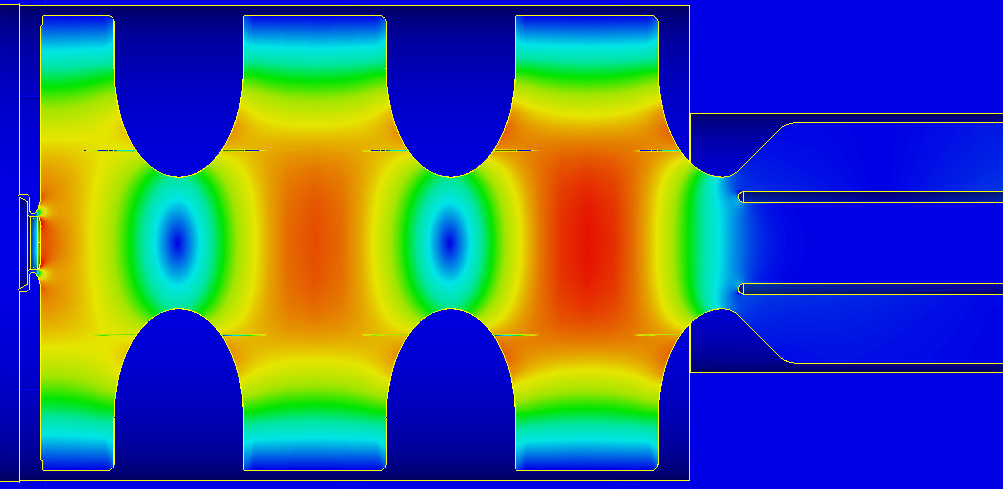} \hfill 

\includegraphics[width=0.4\textwidth]{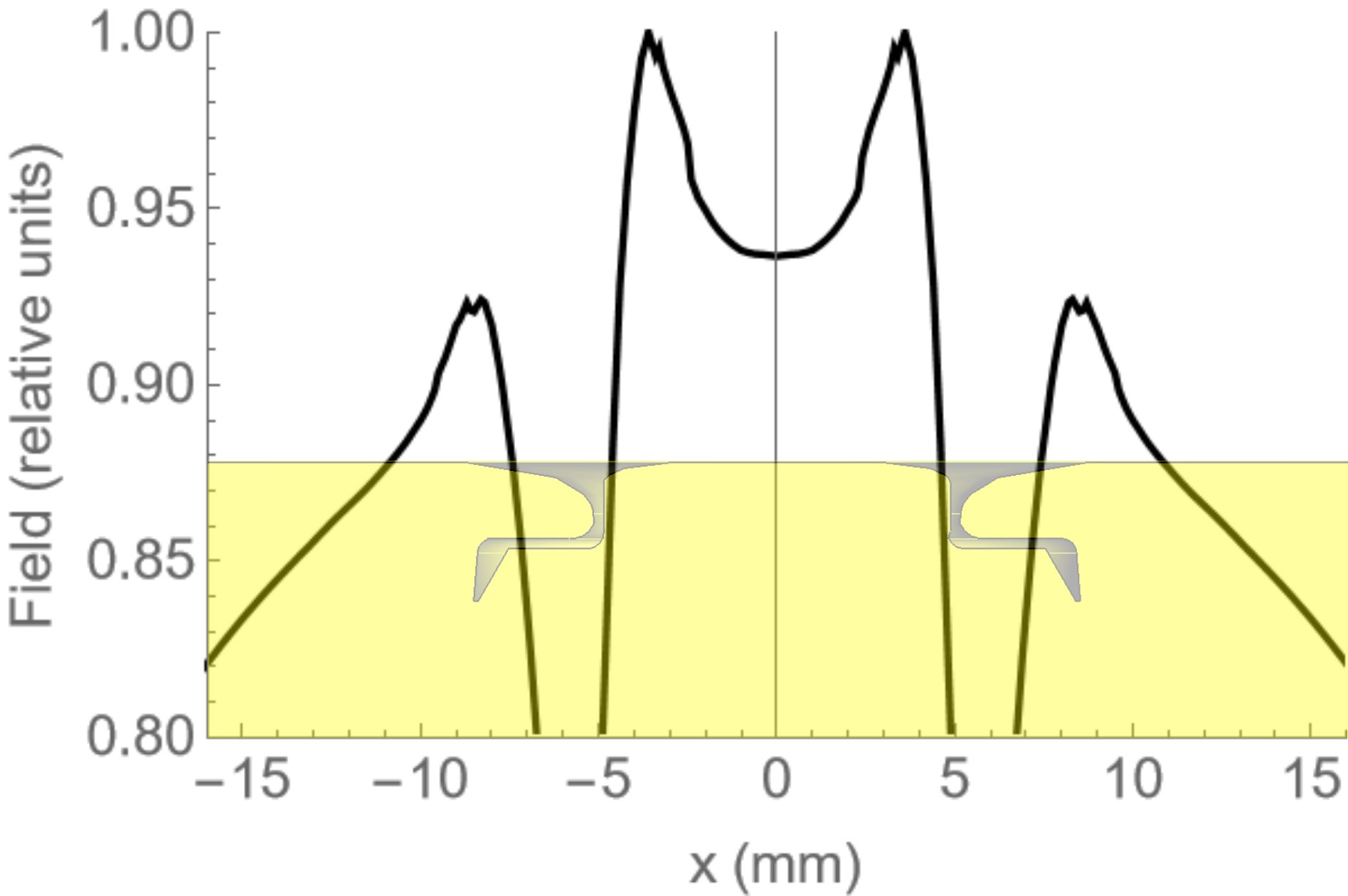} 
\caption{\label{fig:cst2} CLARA gun electric field strength in x-z plane. Top: expanded view of gun, colour denotes the field strength  (red strongest, blue weakest). Bottom: Electric field strength along cathode backplate surface, with cathode region profile superimposed.}
\end{figure}

The gun EM field was simulated using the FEM field solver of CST, using the gun and coupler section of the model. Fig. \ref{fig:cst2} shows the solved electric field of the gun in the x-z plane. To simulate dark current production and transport the PIC simulation of CST is used, which incorporates the EM field generated from the previous step, and the combined gun solenoid and bucking solenoid field modelled by the OPERA package. The field emission from the gun surfaces was modelled by a simplified version of the Fowler Nordheim (FN) formalism, of which a key feature is a strong sensitivity of the emitted charge to the size of the electric field (a description of the FN model can be found at, for example [\onlinecite{PhysRevLett.109.204802}]). The spatial density of the field emission points was uniform over the gun surfaces, and the emitted particles were then tracked through the gun and solenoid fields to the position of the first screen.  

There are several features related to the discretisation of the FEM nature of this model. A tetrahedral mesh is used solve the EM field, which better describes curved surfaces compared to a hexahededral mesh; however the PIC simulation uses a hexahedral mesh to track the particles through the field. The particle generation uses triangulation to model emission points over the 3-D surfaces of the gun. The parameters of each discretisation, and compatibility between them, may affect both the fine and coarse results of the simulation. In the simulations of the simpler pre-upgrade gun~[\onlinecite{jackson:ipac17-wepab088}], convergence of the amount of charge emitted for each surface was achieved by choosing a sufficiently dense set of mesh parameters. The more complex geometries of the upgraded gun did not allow convergence to be achieved; the amount of charge emitted particularly from the curved surfaces of the cathode region continued to vary as the mesh density was increased. The mesh density was increased up to the point where simulation time became impractical, and at this stage the emitted charge from the cathode rim still varied by around 20 \% for a 10 \% increase in mesh density.

\begin{figure}[!htb]
\centering
\includegraphics[width=0.5\textwidth]{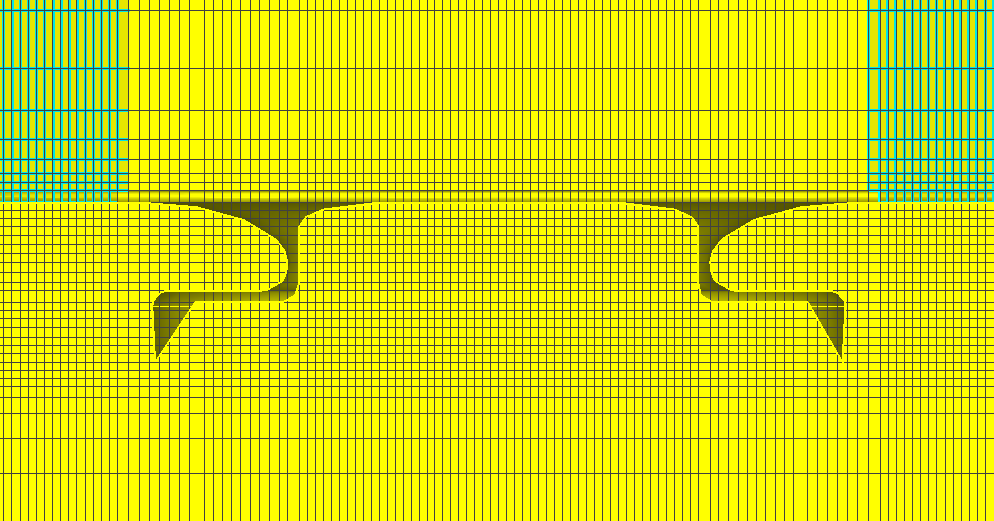}
\caption{\label{fig:cst3} The PIC meshing density in the region of the photo-cathode. The smallest meash size shown here is 0.2 mm}
\end{figure}

The goal of the studies was to examine the appearance of the dark current on the screen, and so convergence of the total emitted and transported charge was not necessary, as long as the parameterisation was sufficent to achieve physically reasonable results (i.e. demonstrating sufficient symmetry and uniformity). Thus the meshing parameters were adjusted to achieve this. Eventually, {\em sub-meshing} was employed; the PIC mesh had different density in different regions of the gun, depending on the geometric complexity of each region. Fig  ~\ref{fig:cst3} shows a closer detail of the meshing density in the area of the cathode where the most complex geometry in the model arises. The smallest mesh size in the transverse and longitudinal dimensions is 0.2 mm. 

The imported OPERA solenoid field has a step size of 2 mm in all three dimensions, and was generated with at mesh size in the beam region (a 20mm radius, 350mm long cylinder) of 2.25mm and larger mesh elsewhere. In making the solenoid field calculations, the `integration' method was used (rather than `nodal'), ensuring that contributions from conductors and magnetised elements are considered from the whole model when calculating the field at a point. This ensures high accuracy, and the calculated solenoid field values did not change when the mesh sizes were varied.

\section{\label{sec:level1}RESULTS AND DISCUSSION}

Following development of the CST model of the upgraded CLARA gun, several experimental observations of dark current were made during the commissioning of the gun, and an attempt was made to match these to the CST simulation. Several different photocathodes of different materials and preparations were used in the commissioning which are detailed in~[\onlinecite{noakes:ipac2023-tupa032}].

Previous general characteristics of the dark current in the pre-upgrade CLARA gun, such as 6D phase space distributions, have been reported~[\onlinecite{jackson:ipac17-wepab088}], and similar general characteristics were seen here. This included the finding that, for nominal operating parameters and assuming uniform emission properties over all surfaces, only the backplate and cathode regions contributed to dark current escaping the gun. With the gun electric field being highest on the cathode and the rims (see Fig.\ref{fig:cst2}) it was the starting point for these studies that emission from the backplate (excluding the backplate rim) would be neglected. Other characteristics included a large energy spread of the dark current from each surface, with significant dark current at the nominal energy of the beam. There was not enough difference in the energy distributions from each surface to suggest the contributions from the flat cathode region and the rims could be separated and identified by dipole spectroscopy. Measurement and monitoring of the amount of dark current emitted from the gun using the WCM was performed routinely throughout commissioning of the upgraded gun as described in~[\onlinecite{noakes:ipac2023-tupa032}]. 

\begin{figure}[!htb]
\centering
\includegraphics[width=0.5\textwidth]{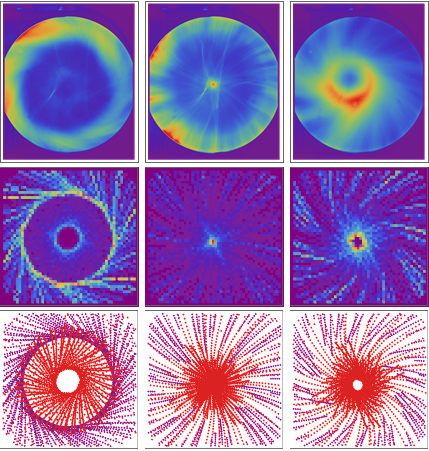}
\caption{\label{fig:results1} Dark current at the first CLARA diagnostic screen for different gun solenoid configurations (represented by columns) for measured and simulated data (represented by rows). Top row: measured. Middle row: simulation including emission from cathode flat surfaces and cathode and backplates rim (colour scale is the particle density). Bottom row: simulation including emission from cathode rim (red) and backplate rim (purple). The columns are three different configurations of solenoid strengths (see text for description) with the middle column the case of the nominal setting for beam focussing, where the solenoid field is cancelled on the cathode surface.}
\end{figure}

The main experimental observations described here concern the appearance of dark current on a scintillating screen near the gun for different solenoid strengths and are shown in Fig. \ref{fig:results1}. During operation of the first photocathode used in the commissioning of the CLARA gun (called cathode `\#13' see ~[\onlinecite{noakes:ipac2023-tupa032}] ), the screen 
image was taken at 3 different solenoid/bucking-solenoid configurations, showing distinctively different coarse and fine features of the dark current. The solenoid field configurations included 1) the main solenoid on but the bucking solenoid off 2) the bucking solenoid set to a strength which cancels the main solenoid field on the cathode surface (which is the nominal setting used for beam focussing) 3) the bucking solenoid and main solenoid on but at strength where the field does not cancel on the cathode surface.
The notional gun field (at its nodal maxima) during these measurements was kept constant during this set of measurements and estimated to be 67 MV/m (using the measured RF power to the gun, a previously measured beam momentum vs RF power calibration, and a simplified simulation with a 1-D model of the field). This may omit some realistic features and errors in the actual gun field. Thus in the CST simulations the overall scaling of the gun field was set to a range (approximately 20 \%) around 67 MV/m, and the results (for all three solenoid values) which gave the closest match to observed screen images are shown here. In a similar way the CST PIC FN model parameters for each surface were adjusted in an ad-hoc manner and a further scaling applied to the emission from each surface to find a closer agreement with the data. The scaling is consistent across the three solenoid configurations. 

 For the simulated images shown in Fig. \ref{fig:results1}, the colour map (middle row) used is the charge per 2D bin, to compare with the intensity colour map of the screen images. The comparison of similar features in screen measurements and simulations shown in Fig. \ref{fig:results1} indicate that dark current originating from field emission from the rims in the cathode region give rise to the distinctive `ring' features. This is clarified by the colour coded comparison of the separated contributions from only the rims (Fig. \ref{fig:results1} bottom row). In the first solenoid configuration this leads to a large bright ring originating from the backplate rim, and a dimmer smaller ring originating from the cathode rim. In the second solenoid configuration the bright central spot originates from the cathode rim with a larger `halo' originating from the backplate rim. In the final solenoid configuration a bright ring of originates from the cathode rim. There may be several factors involved in the relative brightness of rings due to the entire transport of the dark current from source to screen, but more intense emission from the rim surfaces compared to the flat central cathode surface is expected due to electric field enhancement. The surface roughness on the rims is expected to be greater than the flat cathode surface, thus field enhancement on the rims would generate more field emission (a more complete formulation of the FN model [\onlinecite{PhysRevLett.109.204802}] includes parameterisation of roughness in a field enhancement factor $\beta$, and predicts a strong dependence of the amount of emission on surface quality). In addition, the non-enhanced electric field strength varies over the cathode region and reaches maxima on the cathode rim (see Fig. \ref{fig:cst2}.) which again implies an increased contribution to dark current from the cathode rim.


The results indicate that relatively bright central features (middle column of Fig. \ref{fig:results1}) are not necessarily due to field emission from the centre of the cathode. They may instead be field emission from the rims which are then focused to the centre of the screen by the solenoid. Some fine features seen in the screen images may be a result of non-uniformly distributed imperfections on the surfaces, which are not reproduced by the uniformly distributed emission points of the simulations. 

\begin{figure}[!htb]
\centering
\includegraphics[width=0.5\textwidth]{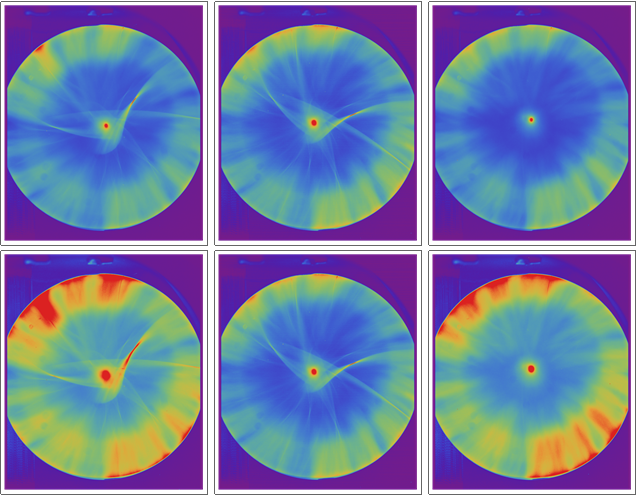}
\caption{\label{fig:results2} Dark current screen images at different stages of CLARA gun conditioning, with different cathode status. Hybrid cathode (`\#7') first and second insertion (left and middle), and diamond turned cathode (`\#16') (right). In the top row the colour scale is adjusted to emphasise the comparison of the features; in the bottom row the color scale is identical for each figure. The similarity of the outer features in the three images indicate emission from the backplate rim, while the difference in the inner features indicate emission from the cathode.}
\end{figure}

There were other further observations made throughout the subsequent gun commissioning and operation of different photocathodes which support these findings. For example during the commissioning of a Mo/Cu hybrid photocathode (named `\#7') heat cleaning was performed, involving retraction and reinsertion; later another hybrid cathode (named `\#16') with different surface preparation (diamond-turned) was inserted. The appearance of dark current on the screen during these changes was observed and compared. This is shown in Fig. ~\ref{fig:results2} where approximately the same gun field strength and solenoid strength was used in all the images (note these approximately matched, but were slightly different to, those used in the middle column of Fig. ~\ref{fig:results1}, with the solenoid field cancelled on the cathode surface). The images taken before and after retraction and reinsertion of the cathode highlight the dark current features from emitters on the cathode compared to those on the surrounding backplate and rim. Similarly the inserting of a different cathode shows a clear difference in some features and similarity on others, again allowing the identification of different sources and confirming the ability of the simulations to predict these features.    
An interesting observation is the difference in brightness of the dark current 
 on the screen for different cathodes and/or different insertions of the same cathode. This may be due to the effect of very small ($\mu$m) differences in positioning of the cathode and/or the length of the different cathodes. The geometrical change this makes to the gun cavity may change the EM field strength on the cathode surfaces 
(as seen in~[\onlinecite{SHU2021165546}]), 
and thus change the intensity of the dark current emission. Some observations of gun tuning temperature during the commissioning suggested the cathode was recessed by a small fraction of a mm at some stages, and the CST field solver estimates significantly higher electric field strength across cathode and backplate surfaces results from this (although complete PIC simulations have not been performed).    
A wall current monitor (WCM) located approximately 20 cm upstream of the screen was used to measure the dark current, which confirms the reduction in dark current (around 15 \%) seen after retraction and reinsertion of the cathode seen in Fig ~\ref{fig:results2}.

\section{\label{sec:level1}CONCLUSIONS}
Despite the lack of quantitative convergence, the results above demonstrate the usefulness of simulations in the ability to characterise the effect of each emitting source on the screen pattern. In the screen images of the dark current the emission from the backplate rim and from the cathode insertion has been identified. This is clear due to the comparison of different experimental conditions (retraction and reinsertion of cathodes), and is also supported by the simulations, where different emission sources can be isolated and the simulated screen features (rings, bright and dark regions) match the measured features in size and shape. For the CLARA gun, these observations lead to the following conclusions.

Field emission from the curved section of the backplate is significant and, for certain parameters, may dominate the dark current transported through the first part of the beam-line. For the nominal beam setting of the solenoids it appears as an outer bright ring on the screen.   

Field emission from the cathode (flat surface and the curved rim) leads to features observed in the central part of the screen, when the solenoid strength is set approximately to its nominal beam focussing setting. The central bright spot may not be due to field emitters in the centre of the cathode, since simulations indicate field emission from the edge of the cathode may be focused to the centre of the screen.  

There are several issues and uncertainties with the simulations that would require further work to evaluate. Our hypothesis on the effect of the cathode position or length on the dark current brightness has not yet been simulated in full. Futhermore there are various ways in which the electric and magnetic fields in the simulations may differ from reality. At the emission surfaces where the particle energy is low this may significantly affect the dynamics of the field emission. In simulation, meshing artefacts may affect the gun electromagnetic field shape, even for perfectly described geometries. Other effects which might be thought small may also be important; for example mutual coil excitation included in the modelling of combined solenoid and bucking solenoid field were observed to make a significant difference to the simulated screen patterns. We do not have detailed in-situ 3-D measurements of the magnetic and electric fields, which may include other such small effects, to compare against the modelled fields.

\section{\label{sec:level1}ACKNOWLEDGEMENTS}

We wish to thank members of ASTeC (Accelerator Science and Technology Centre) and Technology Departments of STFC (Science and Technology Facilities Council) for providing technical and operational support of the CLARA accelerator, which made these experiments possible. We wish to particularly thank I. Gessey and P. Tipping for development of the CST field emission simulations of the CLARA gun.  

\section{\label{sec:level1}REFERENCES}


\bibliography{pub}

\end{document}